\documentstyle[twoside]{article}

\def\Q#1{Q_{#1}}
\def\dc#1{d^{c}_{#1}}
\def\uc#1{u^{c}_{#1}}

\def\L#1{L_{#1}}
\def\ec#1{e^{c}_{#1}}
\def\Nc#1{N^{c}_{#1}}

\def\Tr{{\rm Tr}\, }

\def\eps{\epsilon}

\def\NPB#1#2#3{{\bibit Nucl.\ Phys.}\/ {\bibbf B#1} (#2) #3}
\def\woNPB{{\bibit Nucl.\ Phys.}\/ {\bibbf B}}

\def\PLB#1#2#3{{\bibit Phys.\ Lett.}\/ {\bibbf B#1} (#2) #3}
\def\PRD#1#2#3{{\bibit Phys.\ Rev.}\/ {\bibbf D#1} (#2) #3}

\def\MPLA#1#2#3{{\bibit Mod.\ Phys.\ Lett.}\/ {\bibbf A#1} (#2) #3}

\def\woIJMPA{{\bibit Int.\ J.\ Mod.\ Phys.}\/ {\bibbf A}}

\def\inbar{\,\vrule height1.5ex width.4pt depth0pt}

\def\IC{\relax\hbox{$\inbar\kern-.3em{\rm C}$}}
\def\IQ{\relax\hbox{$\inbar\kern-.3em{\rm Q}$}}
\def\IR{\relax{\rm I\kern-.18em R}}
 \font\cmss=cmss10 \font\cmsss=cmss10 at 7pt

\def\IZ{\relax\ifmmode\mathchoice
 {\hbox{\cmss Z\kern-.4em Z}}{\hbox{\cmss Z\kern-.4em Z}}
 {\lower.9pt\hbox{\cmsss Z\kern-.4em Z}}
 {\lower1.2pt\hbox{\cmsss Z\kern-.4em Z}}\else{\cmss Z\kern-.4em Z}\fi}

\catcode`\@=11
\long\def\@makefntext#1{
\protect\noindent \hbox to 3.2pt {\hskip-.9pt  
$^{{\eightrm\@thefnmark}}$\hfil}#1\hfill}		

\def\thefootnote{\fnsymbol{footnote}}
\def\@makefnmark{\hbox to 0pt{$^{\@thefnmark}$\hss}}	
	
\def\ps@myheadings{\let\@mkboth\@gobbletwo
\def\@oddhead{\hbox{}
\rightmark\hfil\eightrm\thepage}   
\def\@oddfoot{}\def\@evenhead{\eightrm\thepage\hfil
\leftmark\hbox{}}\def\@evenfoot{}
\def\sectionmark##1{}\def\subsectionmark##1{}}

\oddsidemargin=\evensidemargin
\renewcommand{\thefootnote}{\fnsymbol{footnote}}

\newcounter{sectionc}\newcounter{subsectionc}\newcounter{subsubsectionc}
\renewcommand{\section}[1] {\vspace{12pt}\addtocounter{sectionc}{1} 
\setcounter{subsectionc}{0}\setcounter{subsubsectionc}{0}\noindent 
	{\tenbf\thesectionc. #1}\par\vspace{5pt}}
\renewcommand{\subsection}[1] {\vspace{12pt}\addtocounter{subsectionc}{1} 
	\setcounter{subsubsectionc}{0}\noindent 
	{\bf\thesectionc.\thesubsectionc. {\kern1pt \bfit #1}}\par\vspace{5pt}}
\renewcommand{\subsubsection}[1] {\vspace{12pt}\addtocounter{subsubsectionc}{1}
	\noindent{\tenrm\thesectionc.\thesubsectionc.\thesubsubsectionc.
	{\kern1pt \tenit #1}}\par\vspace{5pt}}
\newcommand{\nonumsection}[1] {\vspace{12pt}\noindent{\tenbf #1}
	\par\vspace{5pt}}

\newcounter{appendixc}
\newcounter{subappendixc}[appendixc]
\newcounter{subsubappendixc}[subappendixc]
\renewcommand{\thesubappendixc}{\Alph{appendixc}.\arabic{subappendixc}}
\renewcommand{\thesubsubappendixc}
	{\Alph{appendixc}.\arabic{subappendixc}.\arabic{subsubappendixc}}

\renewcommand{\appendix}[1] {\vspace{12pt}
        \refstepcounter{appendixc}
        \setcounter{figure}{0}
        \setcounter{table}{0}
        \setcounter{lemma}{0}
        \setcounter{theorem}{0}
        \setcounter{corollary}{0}
        \setcounter{definition}{0}
        \setcounter{equation}{0}
        \renewcommand{\thefigure}{\Alph{appendixc}.\arabic{figure}}
        \renewcommand{\thetable}{\Alph{appendixc}.\arabic{table}}
        \renewcommand{\theappendixc}{\Alph{appendixc}}
        \renewcommand{\thelemma}{\Alph{appendixc}.\arabic{lemma}}
        \renewcommand{\thetheorem}{\Alph{appendixc}.\arabic{theorem}}
        \renewcommand{\thedefinition}{\Alph{appendixc}.\arabic{definition}}
        \renewcommand{\thecorollary}{\Alph{appendixc}.\arabic{corollary}}
        \renewcommand{\theequation}{\Alph{appendixc}.\arabic{equation}}
        \noindent{\tenbf Appendix \theappendixc #1}\par\vspace{5pt}}
\newcommand{\subappendix}[1] {\vspace{12pt}
        \refstepcounter{subappendixc}
        \noindent{\bf Appendix \thesubappendixc. {\kern1pt \bfit #1}}
	\par\vspace{5pt}}
\newcommand{\subsubappendix}[1] {\vspace{12pt}
        \refstepcounter{subsubappendixc}
        \noindent{\rm Appendix \thesubsubappendixc. {\kern1pt \tenit #1}}
	\par\vspace{5pt}}

\newcommand{\textlineskip}{\baselineskip=13pt}
\newcommand{\smalllineskip}{\baselineskip=10pt}

\def\eightcirc{
\begin{picture}(0,0)
\put(4.4,1.8){\circle{6.5}}
\end{picture}}
\def\eightcopyright{\eightcirc\kern2.7pt\hbox{\eightrm c}}

\def\abstracts#1#2#3{{
	\centering{\begin{minipage}{4.5in}\baselineskip=10pt\footnotesize
	\parindent=0pt #1\par 
	\parindent=15pt #2\par
	\parindent=15pt #3
	\end{minipage}}\par}} 

\newcommand{\bibit}{\nineit}
\newcommand{\bibbf}{\ninebf}
\renewenvironment{thebibliography}[1]
	{\frenchspacing
	 \ninerm\baselineskip=11pt
	 \begin{list}{\arabic{enumi}.}
	{\usecounter{enumi}\setlength{\parsep}{0pt}
	 \setlength{\leftmargin 12.7pt}{\rightmargin 0pt} 
	 \setlength{\itemsep}{0pt} \settowidth
	{\labelwidth}{#1.}\sloppy}}{\end{list}}

\newcounter{itemlistc}
\newcounter{romanlistc}
\newcounter{alphlistc}
\newcounter{arabiclistc}

\newcommand{\fcaption}[1]{
        \refstepcounter{figure}
        \setbox\@tempboxa = \hbox{\footnotesize Fig.~\thefigure. #1}
        \ifdim \wd\@tempboxa > 5in
           {\begin{center}
        \parbox{5in}{\footnotesize\smalllineskip Fig.~\thefigure. #1}
            \end{center}}
        \else
             {\begin{center}
             {\footnotesize Fig.~\thefigure. #1}
              \end{center}}
        \fi}

\newcommand{\tcaption}[1]{
        \refstepcounter{table}
        \setbox\@tempboxa = \hbox{\footnotesize Table~\thetable. #1}
        \ifdim \wd\@tempboxa > 5in
           {\begin{center}
        \parbox{5in}{\footnotesize\smalllineskip Table~\thetable. #1}
            \end{center}}
        \else
             {\begin{center}
             {\footnotesize Table~\thetable. #1}
              \end{center}}
        \fi}

\def\@citex[#1]#2{\if@filesw\immediate\write\@auxout
	{\string\citation{#2}}\fi
\def\@citea{}\@cite{\@for\@citeb:=#2\do
	{\@citea\def\@citea{,}\@ifundefined
	{b@\@citeb}{{\bf ?}\@warning
	{Citation `\@citeb' on page \thepage \space undefined}}
	{\csname b@\@citeb\endcsname}}}{#1}}

\newif\if@cghi
\def\cite{\@cghitrue\@ifnextchar [{\@tempswatrue
	\@citex}{\@tempswafalse\@citex[]}}
\def\citelow{\@cghifalse\@ifnextchar [{\@tempswatrue
	\@citex}{\@tempswafalse\@citex[]}}
\def\@cite#1#2{{$\null^{#1}$\if@tempswa\typeout
	{IJCGA warning: optional citation argument 
	ignored: `#2'} \fi}}
\newcommand{\citeup}{\cite}

\def\pmb#1{\setbox0=\hbox{#1}
	\kern-.025em\copy0\kern-\wd0
	\kern.05em\copy0\kern-\wd0
	\kern-.025em\raise.0433em\box0}

\def\fnt#1#2{\footnotetext{\kern-.3em
	{$^{\mbox{\scriptsize #1}}$}{#2}}}

\def\fpage#1{\begingroup
\voffset=.3in
\thispagestyle{empty}\begin{table}[b]\centerline{\footnotesize #1}
	\end{table}\endgroup}

\def\runninghead#1#2{\pagestyle{myheadings}
\markboth{{\protect\footnotesize\it{\quad #1}}\hfill}
{\hfill{\protect\footnotesize\it{#2\quad}}}}
\font\tenrm=cmr10
\font\tenit=cmti10 
\font\tenbf=cmbx10
\font\bfit=cmbxti10 at 10pt
\font\ninerm=cmr9
\font\nineit=cmti9
\font\ninebf=cmbx9
\font\eightrm=cmr8

\def\qed{\hbox{${\vcenter{\vbox{			
   \hrule height 0.4pt\hbox{\vrule width 0.4pt height 6pt
   \kern5pt\vrule width 0.4pt}\hrule height 0.4pt}}}$}}

\renewcommand{\thefootnote}{\fnsymbol{footnote}}	

\begin{document}

\runninghead{Minimal Supersymmetric Standard Model
Manuscripts $\ldots$} {Minimal Supersymmetric Standard Model
Manuscripts $\ldots$}

\normalsize\textlineskip
\thispagestyle{empty}
\setcounter{page}{1}


\vspace*{0.88truein}
\rightline{ACT-15/00}
\rightline{CTP-TAMU-27/00}
\rightline{hep-ph/0011020}
\rightline{November 2000}
\rightline{\phantom{November 2000}}

\fpage{1}
\centerline{\bf MINIMAL SUPERSTRING STANDARD MODEL:}
\vspace*{0.035truein}
\centerline{\bf A REVIEW\footnote{Talk presented at DPF 2000, 
Columbus, Ohio, August 9-12, 2000.}}
\vspace*{0.37truein}
\centerline{\footnotesize Gerald B.\ Cleaver}
\vspace*{0.015truein}
\centerline{\footnotesize\it 
Center for Theoretical Physics,
Department of Physics}
\baselineskip=10pt
\centerline{\footnotesize\it Texas A \& M University, 
College Station, Texas, 77843, USA}
\centerline{\footnotesize\it and}
\centerline{\footnotesize\it 
Astro Particle Physics Group} 
\centerline{\footnotesize\it 
Houston Advanced Research Center (HARC)} 
\baselineskip=10pt
\centerline{\footnotesize\it 
Woodlands, Texas, 77381 USA}

\vspace*{0.21truein}
\abstracts{I review a heterotic--string solution
in which the observable sector effective field theory just below 
the string scale reduces to that of the MSSM, 
with the standard observable gauge group being just
$SU(3)_C\times SU(2)_L\times U(1)_Y$ and the
$SU(3)_C\times SU(2)_L\times U(1)_Y$--charged 
spectrum of the observable sector consisting solely of the MSSM spectrum.
Associated with this model is a set of distinct flat directions of 
vacuum expectation values (VEVs) of fields
that all produce solely the MSSM spectrum.
Some of these directions only involve VEVs of non--Abelian singlet fields
while others also contain VEVs of non--Abelian charged fields.
The phenomenology of theses flat directions is summarized.}{}{}



\textheight=7.8truein
\setcounter{footnote}{0}
\renewcommand{\thefootnote}{\alph{footnote}}

\section{Minimal Superstring Standard Model}
\noindent
The most realistic string models found to date have been constructed
in the free fermionic formulation\citeup{fff} of the heterotic--string.
A large number of three generation models, which differ in their
detailed phenomenological characteristics, have been built.\citeup{real}
All these models share an underlying $\IZ_2\times \IZ_2$ orbifold 
structure, which naturally gives rise to three generations
with the $SO(10)$ embedding of the Standard Model spectrum.\citeup{nahe} 
In some of these models the observable 
sector gauge group directly below the string scale is a Grand Unified Theory, 
while in others it is the (MS)SM gauge group, 
$SU(3)_C\times SU(2)_L\times U(1)_Y$, joined by a few extra $U(1)$ symmetries.
In chiral three generation models of the latter class, one of the additional 
Abelian gauge groups is inevitably anomalous. That is, the trace of the
$U(1)_A$ charge, $\Tr Q^{(A)}\ne 0$, generates a Fayet--Iliopoulos (FI) term, 
$\eps\equiv (g^2_s M_P^2/192\pi^2)\Tr Q^{(A)}$.
The FI term
breaks supersymmetry near the Planck scale, and destabilizes
the string vacuum. Supersymmetry is restored and the vacuum is
stabilized if there exists a direction, ${\phi}=\sum_i\alpha_i \phi_i$,
in the scalar potential which is 
$F$--flat to sufficient order
(general arguments suggest this means to at least 17$^{\rm th}$ order),  
and also
$D$--flat with respect to all non--anomalous gauge symmetries,
and in which $\sum_i Q_i^{(A)}\vert\alpha_i\vert^2$ and $\eps$
are of opposite sign.
If such a direction exists it will acquire a VEV
cancelling the anomalous $D$--term, restoring supersymmetry and
stabilizing the string vacuum. 

In addition to possessing an anomalous $U(1)_A$ symmetry,
chiral three generation   
$SU(3)_C\times SU(2)_L\times U(1)_Y\times \prod_i U(1)_i$ 
string models generically contain numerous non--MSSM  
$SU(3)_C\times SU(2)_L\times U(1)_Y$--charged states,
some of which only carry  
MSSM$\times \prod_i U(1)_i$ 
charges and others of which also carry hidden sector (non)--Abelian
charges. Recently, we showed that in some of these models 
it is actually possible to decouple all such non--MSSM states from the low
energy effective field theory. 
For example, in the ``FNY'' model first presented in Ref.\ 4,
we discovered there are
several flat directions\citeup{cfn1,cfn2} for which almost all
MSSM--charged exotics receive FI masses
(typically of the scale $5\times 10^{16}$ GeV to $1\times 10^{17}$ GeV)
while the remaining MSSM--charged exotics (usually composed of simply a
$SU(3)_C$ triplet/anti--triplet pair) acquire masses slightly suppressed 
below the FI scale by a factor of ${\cal O}(\frac{1}{10}$ to $\frac{1}{100})$.
Some of our flat directions accomplishing this feat contain only 
VEVs of non--Abelian singlet fields\citeup{cfn2,cfn3}
while others of ours also contain VEVs of non--Abelian charged 
fields\citeup{cfn4,cfn5}. Along these directions,
exactly three generations of 
($\Q{i}$, $\uc{i}$, $\dc{i}$, $\L{i}$, $\ec{i}$, $\Nc{i}$)
fields and a pair of electroweak Higgs, $h_u$ and $h_d$, are the only
MSSM--charged fields that remain massless significantly below the FI scale. 
The non--MSSM--charged singlet fields 
and hidden sector non--Abelian fields that also 
remain massless below the FI scale vary with the flat direction chosen. 

\section{Flat Direction Phenomenology}
\noindent

The complete massless spectrums and superpotentials through 
sixth order resulting from the anomaly-cancelling
singlet and non-Abelian flat directions appear, respectively, in 
Refs.\ 7 and 10.
In all of the singlet directions 
and in many of the non-Abelian flat directions, 
the leading mass terms were found to be
$Q_1u_1^c{\bar h}_1$ and $Q_3d_3^ch_3$.  
In other words, 
the left--handed components of the heaviest up-like and down-like
quarks live in different multiplets. 
Thus, these directions possess the phenomenological problem of  
effectively interchanging the strange and bottom masses.
Importantly, several non-Abelian flat directions do not present this difficulty
and appear to offer much more viable phenomenology.

We have discovered that 
a few extra Abelian symmetries usually survive along
both singlet and non-Abelian flat directions. 
Generically, these extra local $U(1)_i$  
could not have been embedded in $SO(10)$ or $E_6$ 
GUTS.\citeup{cfn5} 
Relatedly, a common feature in the surviving $U(1)_i$ combinations 
is a flavor non--universality.
Thus, the distinctive collider signatures
of a $Z^\prime$ arising from one such symmetry
will be a non--universality in the production of the different
generations. An additional $Z^\prime$ of this type 
has also been suggested as playing a role
in suppressing proton decay in supersymmetric extensions
of the Standard Model.\citeup{pati}

Another very interesting aspect of our non-Abelian flat directions
is that the related Higgs mass eigenstates, $h_u$ and $h_d$, 
each contain several components, with the weights of the components
often varying by several orders of magnitude.
Furthermore, different quark and lepton generations
couple to different Higgs components. Thus, even when all three 
generations of quark and lepton mass terms result from low 
order superpotential terms, a natural generational mass hierarchy
appears. This could provide a novel explanation for the 
approximate $10^{-5}: 10^{-3}: 1$ generational mass ratio.

In Ref.\ 7
we presented several reasons why 
non-Abelian VEVs are, ideed, likely required 
for a phenomenologically viable
low energy effective MSSM, at least for the FNY string model.
Evidence has also been presented in the past suggesting this might be true 
as well for all string-derived MSSM 
$\IZ_2\times \IZ_2$ models. This implies that there 
is truly significant worth in exploring 
the generic properties of 
non-Abelian flat directions in $\IZ_2\times \IZ_2$ models  
that contain exactly the MSSM three generations and two Higgs doublets
as the only MSSM--charged fields in the low energy effective field theory.

\nonumsection{Acknowledgements}
\noindent

G.C.~thanks the organizers of DPF 2000 for an excellent 
educational and enjoyable conference. 
The work discussed herein was done in collaboration
with A.E.~Faraggi, D.V.~Nanopoulos, \& J.W.~Walker.
This work is supported in part by  
DOE Grants DE--FG--0395ER40917 (GC,DVN,JW)
and DE--FG--0294ER40823 (AF).

\nonumsection{References}

\end{document}